\documentclass[prl,groupaddress,twocolumn,showpacs]{revtex4}
\usepackage{epsfig,amsmath,amssymb,graphics,color,calc}

\newcommand{\be}{\begin{equation}}
\newcommand{\ee}{\end{equation}}
\newcommand{\ba}{\begin{eqnarray}}
\newcommand{\ea}{\end{eqnarray}}

\begin{document}

\title{Efficient measurement of linear susceptibilities in
molecular \\ simulations: Application to aging supercooled liquids}

\author{Ludovic Berthier}
\affiliation{Laboratoire des Collo{\"\i}des, Verres
et Nanomat{\'e}riaux, UMR 5587, Universit{\'e} Montpellier II and CNRS,
34095 Montpellier, France}

\date{\today}

\begin{abstract}
We propose a new method to measure time-dependent linear  
susceptibilities in molecular simulations, which does not require the use of
nonequilibrium simulations, subtraction
techniques, or fluctuation-dissipation theorems.
The main idea is an exact reformulation of linearly 
perturbed quantities in terms of observables
accessible in unperturbed trajectories. We have applied these ideas
to two supercooled liquids in their nonequilibrium aging regime. 
We show that previous work had underestimated deviations from 
fluctuation-dissipation relations in the case of a Lennard-Jones 
system, while our results for silica are in 
qualitative disagreement with earlier results.
\end{abstract}



\maketitle

Correlation and response functions play a major role in 
condensed matter physics as they directly probe static 
and dynamic properties at a microscopic level~\cite{forster}. 
At thermal equilibrium, linear response theory permits
the derivation of fluctuation-dissipation relations
between conjugated susceptibilities and correlations, so that
both types of measurements become equivalent~\cite{hansen}. 
Depending on the technique used, experiments or simulations
access one or the other quantity. For liquids, neutron scattering 
experiments will for instance be sensitive to spontaneous 
fluctuations of the density, while dielectric spectroscopy 
detects the response induced by an electric 
field~\cite{hansen}. 
Numerical simulations mainly focus on spontaneous fluctuations 
and probe microscopic dynamics via
correlation functions~\cite{allen}. 
However, there exist cases where the numerical
measurement of response functions becomes necessary, for instance 
when correlation functions become too noisy to be detected~\cite{cross}, 
or in nonequilibrium situations, where correlation and response functions  
contain distinct physical information because fluctuation-dissipation
theorems (FDT) do not hold~\cite{jorge}. Quantifying FDT ``violations''
from the simultaneous measurement of correlation and response functions
is an active field of research~\cite{reviewfdt}.
In this work we propose an efficient method to access 
linear response functions in numerical simulations of 
molecular systems. As a physically relevant 
situation we apply this novel technique to 
study response functions of glass-forming liquids 
undergoing physical aging after a sudden quench to low temperature.

Direct measurements of linear susceptibilities usually proceed as 
follows. Consider of system of $N$ particles described by
coordinates, ${\vec r} \equiv \{ {\vec r}_i, i=1,\cdots,N \}$,
momenta, ${\vec p} \equiv \{ {\vec p}_i, i=1,\cdots,N \}$, 
masses $m_i$, and a Hamiltonian ${\cal H}({\vec r}, {\vec p})$
containing a kinetic part, ${\cal K}({\vec p}) = \sum_i 
{{\vec p}_i}^2/(2m_i)$, 
and a potential part, ${\cal V}({\vec r})$.  
We first consider Newtonian dynamics, as used in 
Molecular Dynamics (MD):
\be
{\dot {\vec r}_i} = 
\partial {\cal H}/\partial {\vec p}_i, \quad {\dot {\vec p}_i} 
= - \partial {\cal H} / \partial {\vec q}_i.
\label{newton}
\ee 
Physical observables, $A(t) \equiv A [{\vec p}(t), {\vec r}(t)]$,
can be measured at any time in a simulation, and correlation functions, 
$C(t,t') = \langle A(t) B(t') \rangle_0$, 
are obtained by averaging over repeated measurements. 
The subscript ``$0$'' indicates that averages are performed 
over unperturbed trajectories, and we suppose that 
$\langle A(t) \rangle_0 = 0$. In systems which are 
time-translationally invariant, two-time quantities only depend
on $t-t'$ but we retain the $(t,t')$ notation as we shall 
also study non-stationary systems.

To measure a response function, an external field of 
constant amplitude $h$, conjugated to $B(t)$, 
is introduced at time $t'$, such that the Hamiltonian  
contains the additional term $\delta {\cal H} = - h B$
for $t>t'$. A linear susceptibility can then be defined:
\be
\chi(t,t') = \int_{t'}^t dt'' \frac{\partial \langle A(t) \rangle_h}{\partial 
h(t'')}
\Bigg|_{h \to 0},
\label{chi}
\ee  
Step responses are considered for simplicity 
but the discussion holds more generally.
The average in (\ref{chi}) 
is with the field switched on,  
the zero-field limit comes from repeated
measurements with fields of decreasing amplitude.
In practice, a compromise is sought between large fields
introducing unwanted non-linear effects,
and small fields resulting in poor signals.
Such a non-equilibrium technique suffers from a serious drawback. 
Averages in (\ref{chi}) are taken over perturbed trajectories, so that
susceptibilities can only be measured one at a time, 
contrary to correlation functions which can be simultaneously
measured and time averaged in a single unperturbed trajectory. 

An alternative would be to perform the derivative in 
Eq.~(\ref{chi}) {\it before} taking the average. This is precisely 
how the FDT is derived~\cite{hansen}. Averages are first 
expressed in terms of the distribution function. 
Its thermal equilibrium (Gibbs-Boltzmann) form 
at temperature $T$ is then assumed, 
and the derivative is computed analytically~\cite{hansen}:  
\be 
\chi(t,t') = \frac{1}{T} \left[ C(t,t) - C(t,t') \right], 
\label{fdt}
\ee
where we have set Boltzmann's constant to unity. 
An important and well-known feature of the FDT in Eq.~(\ref{fdt}) 
is that the right 
hand side is evaluated using unperturbed trajectories, 
the temperature prefactor reminding us that thermal equilibrium
is assumed, implying that Eq.~(\ref{fdt}) cannot be used
to measure $\chi(t,t')$ far from equilibrium.

The idea introduced in this paper is to perform the derivative 
before doing the average {\it without} assuming thermal equilibrium.
Similar ideas were recently discussed for discrete 
spins~\cite{ricci}. 
In MD simulations, the subtraction
technique~\cite{ciccotti} is a finite-field approximation of 
this idea: Two simulations are run in parallel starting from the same 
configuration at time $t'$, one with $h=0$, the other with 
a small field, $h$. The susceptibility reads:
$\chi(t,t') \approx \left( \langle A(t)\rangle - \langle
A(t) \rangle_0 \right)/h$. 
Non-equilibrium techniques are in fact 
unnecessary~\cite{ciccotti}, since the $h \to 0$ limit can be taken 
directly from (\ref{newton}) using 
perturbation theory~\cite{hubbard} to devise 
an unperturbed technique.
The quantities ${\vec \chi}_i \equiv 
\partial {\vec r}_i/\partial h$ and 
${\vec \varphi}_i \equiv \partial {\vec p}_i/\partial h$
evolve as~\cite{hubbard}:
\be 
{\dot {\vec \chi}_i} = \frac{ {\vec \varphi}_i }{m_i} - 
\frac{\partial B({\vec r},{\vec p})}{\partial {\vec p}_i}, \quad
{\dot {\vec \varphi}_i} = \frac{\partial  B({\vec r}, {\vec p})}{\partial 
{\vec r}_i}
- \sum_{j=1}^N \frac{\partial^2 {\cal V}({\vec r})}{\partial {\vec r}_i 
\partial {\vec r}_j} \cdot {\vec \chi}_j.   
\label{newton2}
\ee
The susceptibility $\chi(t,t')$ can now be evaluated from 
{\it unperturbed} trajectories: 
\be
\chi(t,t') = \left\langle 
\sum_{i=1}^N \left( 
\frac{\partial A({\vec r},{\vec p})}{\partial {\vec r}_i} \cdot {\vec \chi}_i
+ \frac{\partial A({\vec r},{\vec p})}{\partial 
{\vec p}_i} \cdot {\vec \varphi}_i \right)
\right\rangle_0.
\label{toobad}
\ee

To illustrate the result in Eq.~(\ref{toobad}) we have performed 
MD simulations of a 80:20 binary Lennard-Jones (LJ) 
system composed of $N=10^3$ particles at density $\rho=1.2$.    
Particles interact with a LJ 
potential with parameters that can be found in ~\cite{KA},
chosen to avoid crystallization at low temperature,
and to study the properties of glass-forming liquids.
Technical details of our simulations are as in the original
paper~\cite{KA}. When the temperature gets lower than 
$T \approx 1$ (we use LJ units~\cite{KA}), the dynamics
dramatically slows down, and the system cannot be equilibrated 
in computer simulations below $T\approx 0.43$.  

\begin{figure}
\psfig{file=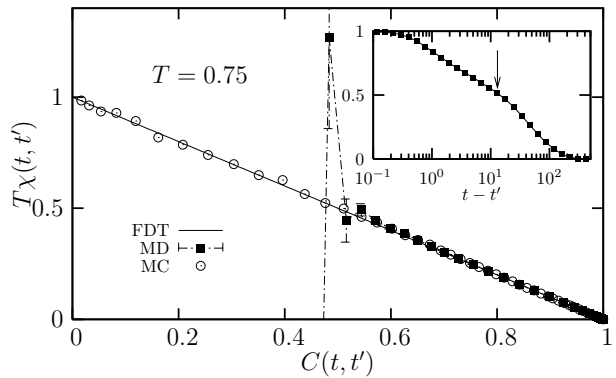,width=8.cm}
\caption{\label{inset}
Simultaneous measurement of susceptibility $\chi(t,t')$ 
and correlation $C(t,t')$ in $10^3$ independent 
unperturbed trajectories  
at $T=0.75$ in the LJ system 
using Eq.~(\ref{toobad}) for MD and Eq.~(\ref{ok}) for MC.
For MD the noise diverges exponentially and $\chi(t,t')$ cannot 
be evaluated for $t-t' > 10$, as indicated in the inset
showing $C(t,t')$ measured in MD.
In MC simulations $\chi(t,t')$ perfectly follows the FDT prediction
indicated by a full line over the whole time range.}
\end{figure}

We perform equilibrium simulations where we simultaneously
solve (\ref{newton})
and (\ref{newton2}) to evaluate $\chi(t,t')$ from (\ref{toobad}), and 
the correlation $C(t,t')$.
We focus on the following observables: 
$A(t) = N^{-1} \sum_j \epsilon_j
\exp [i {\vec k} \cdot {\vec r}_j(t)]$
and $B(t) = 2 \sum_j \epsilon_j \cos [ {\vec k} \cdot {\vec r}_j(t)]$, 
where $\epsilon_j = \pm 1$ is a bimodal random variable of 
mean 0~\cite{hansen}, 
such that $C(t,t')$ corresponds to the
self-intermediate scattering function~\cite{hansen}.
The numerical burden is a mere factor two since one integrates
$12N$ instead of $6N$ equations of motion.
For $T=1.0$, dynamics is fast and 
$\chi(t,t')$ can be evaluated in a few runs, 
as can be checked using the FDT.
For $T=0.75$, where the relaxation time is
$\approx 50$ (see inset of Fig.~\ref{inset}), the fundamental
limitation of the technique appears. 
In Fig.~\ref{inset} we represent $T \chi(t,t')$ evaluated
from $10^3$ independent runs using (\ref{toobad}), 
as a function of $C(t,t')$. FDT predicts
the linear relation shown as a full line. 
For $t-t' \lesssim 5$,
$\chi(t,t')$ follows the FDT. For larger $t-t'$,
the noise in the susceptibility diverges exponentially and no reliable
measurement can be performed, as in 
subtraction techniques. Because the system 
is chaotic, nearby trajectories diverge exponentially quickly: 
While linear response fails at the level of trajectories~\cite{vankampen}, 
it holds at the probabilistic level~\cite{ciccotti}, 
as suggested by the FDT derivation outlined above.

The above exercise suggests that in Monte-Carlo (MC) simulations, 
where phase space is sampled probabilistically rather than
deterministically, response functions could be 
efficiently evaluated. In a standard MC simulation~\cite{allen},
a configuration, ${\cal C}_{t}$, is reached at time $t$. A
trial configuration, ${\cal C}_t'$, is accessed with 
acceptance rate $A_{{\cal C}_t \to {\cal C}_t'}$, generally defined 
from the energy change between the two configurations, 
e.g. the Metropolis rule~\cite{allen} used in the following. The transition
probability from ${\cal C}_t$ to ${\cal C}_{t+1}$ reads:
${\cal W}_{{\cal C}_t \to {\cal C}_{t+1}} = 
\delta_{{\cal C}_{t+1},{\cal C}_t'} A_{{\cal C}_t \to {\cal C}_t'}
+ \delta_{{\cal C}_{t+1},{\cal C}_{t}} (1-A_{{\cal C}_t \to {\cal C}_t'})$.
Averages now mean sampling a large number, ${\cal N}$, of trajectories, 
$\langle A(t) B(t') \rangle_0 = {\cal N}^{-1} \sum_{k=1}^{\cal N}
A_k(t) B_k(t') P_k(t' \to t)$, 
where $A_k(t)$ is the value of 
$A$ at time $t$ in trajectory $k$, and $P_k(t \to t')$ is the 
probability of trajectory $k$ between times $t'$ and $t$ starting
from ${\cal C}_{t'}$, 
$P_k(t' \to t) = \prod_{t''=t'}^{t-1} {\cal W}_{{\cal C}^k_{t''} \to 
{\cal C}^k_{t''+1}}$,
where ${\cal C}^k_{t''}$ is the configuration visited at 
time $t''$ in trajectory $k$.
The susceptibility reads $\chi(t,t') = \partial_h 
[ {\cal N}^{-1} \sum_k A_k(t) P_k(t' \to t) ]$, and
can be reformulated as an {\it unperturbed} average,
\be
\chi(t,t') = \langle A(t) R(t' \to t) \rangle_0, 
\label{ok}
\ee
where $R(t' \to t)  \equiv  
\sum_{t''} \partial_h  \ln ( {\cal W}_{{\cal C}^k_{t''} \to 
{\cal C}^k_{t''+1}})$.
In Fig.~\ref{inset} we report the simultaneous measurement
of $\chi(t,t')$, estimated via (\ref{ok}), and of $C(t,t')$
using $10^3$ independent MC runs of the 
binary Lennard-Jones mixture described above for $T=0.75$. 
(The details of the numerics appeared recently~\cite{tobepisa}.)
The measurement now easily extends over the whole range of timescale 
over which 
$C(t,t')$ changes, and FDT is perfectly obeyed.
Although MC trajectories are chaotic,
no exponential divergence of the noise
is observed, at variance with the MD case.
What Eq.~(\ref{ok})
in fact does is to use a {\it single} unperturbed trajectories 
to evaluate the value the observable $A(t)$ would have taken 
if an infinitesimal field had been applied.
Additionally, the evaluation of Eq.~(\ref{ok})
is computationally free since it only requires 
updating one additional observable, $R (t' \to t)$, 
during the production of unperturbed trajectories.  
Finally, several susceptibilities and correlations
may now be computed during the same simulation, and time
averaging is easily implemented. 
The main limitation of the method is again statistics:
$\chi(t,t')$ now takes the form of a multi-time correlator,
and its measurement becomes statistically costly as $t-t'$ gets too  
large. We find an algebraic growth of the noise, as in spin 
systems~\cite{ricci}, 
which is nevertheless a drastic improvement over exponential growth.
A second drawback is the need to replace Newtonian by Monte-Carlo
dynamics since the resulting {\it dynamics} are not necessarily
equivalent. Quantitative agreement between 
MC and MD dynamics was recently reported for 
the LJ system  described above~\cite{tobepisa}.

We now apply Eq.~(\ref{ok}) to measure $\chi(t,t')$ 
after a sudden quench to very low temperature.
Physical properties of the system now depend
on the time $t'$ spent since the quench, the system ``ages''~\cite{JL}. 
Energy slowly decreases with time, 
while dynamics gets slower~\cite{JL}.
The FDT in Eq.~(\ref{fdt}) no more applies, and 
the following generalization was suggested 
for glassy materials~\cite{teff}, 
\be
\frac{\partial}{\partial t'} \chi(t,t') = - \frac{X(t,t')}{T} 
\frac{\partial}{\partial t'} C(t,t'),
\label{fdr}
\ee
where $X(t,t')$ is the fluctuation-dissipation ratio (FDR), 
$X(t,t')= 1$ at equilibrium. Deviations of the FDR from unity 
serve to quantify the distance from equilibrium~\cite{teff}.

\begin{figure}
\psfig{file=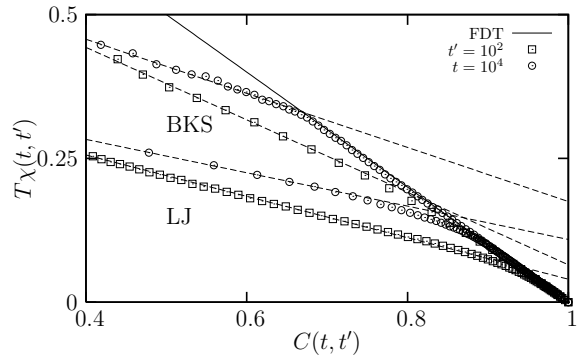,width=7.5cm}
\caption{\label{figttw}
Simultaneous measurement of $\chi(t,t')$ and $C(t,t')$ in aging
LJ ($T=0.4$, $k=6.7$) and silica ($T=2500$~K, $k=2.7$~\AA$^{-1}$). 
Fitting the nonequilibrium part of the FD plots (dashed line)
for fixed-$t$ parametrizations directly yields 
the FDRs $x=0.29$ (LJ) and $x=0.49$ (BKS).
Incorrectly extracting $x$ from fixed-$t'$ data
would yield 0.36 (LJ) and 0.63 (BKS), seriously underestimating 
FDT deviations.}
\end{figure}

Earlier attempts to measure $X(t,t')$ in molecular 
glasses~\cite{fdrsimu,fdrbks}
used the following protocol:
quench the system at $t'=0$; apply a small 
field and measure $\chi(t,t')$ 
for times $t \ge t'$;  build a parametric ``FD plot''
of $\chi(t,t')$ vs $C(t,t')$.
Crucially, this amounts to replacing $\partial_{t'}$ by $\partial_t$ 
in (\ref{fdr}), a procedure which is correct if 
$X(t,t')$ is not an explicit function of $t$ and 
$t'$~\cite{foot}.
Unbiased FDR measurements require instead 
the evaluation of $\chi(t,t')$ at fixed 
time $t$ for different $t'$, so that
the FDR can be graphically deduced from the slope, $-X(t,t')/T$, 
of FD plots. This is numerically too costly
if non-equilibrium techniques are used. 
The difficulty is easily overcome with Eq.~(\ref{ok}), 
and we shall therefore report the first unbiased
FDR measurements in aging molecular liquids.

In Fig.~\ref{figttw} we use both time 
parametrizations to build FD plots in two glass-formers:
the LJ system described above, and the BKS model for 
silica~\cite{bkssimu}. The LJ results are qualitatively consistent
with earlier reports~\cite{fdrsimu}. 
The plots consist of two distinct pieces,
FDT being satisfied for small $t-t'$, 
``violated'' for large $t-t'$. 
Strikingly, FD plots are well-described by two straight lines, leading
to a sensible definition of a constant FDR, $x$, at large $t-t'$.
However, it is obvious in Fig.~\ref{figttw} that 
(incorrectly) estimating $x$ from fixed-$t'$ measurements
yields values that seriously differ from unbiased estimates 
from fixed-$t$ data, an
error made in all previous FDR measurements~\cite{fdrsimu}. 
Both estimates only become equivalent if 
a non-trivial limiting FD plot is found at large time~\cite{teff}.

For silica, we find similar FD plots, 
and similar quantitative discrepancies between both 
time parametrizations. The disagreement with earlier
results is more pronounced for silica since FDR larger than unity 
were reported~\cite{fdrbks}. We have repeated our measurements
at several temperatures between 500 and 2500~K, 
several wavevectors from 0.3 to 13 \AA$^{-1}$, 
both for Si and O atoms. We consistently find 
FD plots as in Fig.~\ref{figttw} with $X(t,t')<1$.
We have numerically checked that this discrepancy cannot
be explained by non-linear effects potentially present in the data
of Ref.~\cite{fdrbks}. Using non-equilibrium techniques 
with large fields we find that non-linear effects yield 
even smaller apparent FDR values. 

\begin{figure}
\psfig{file=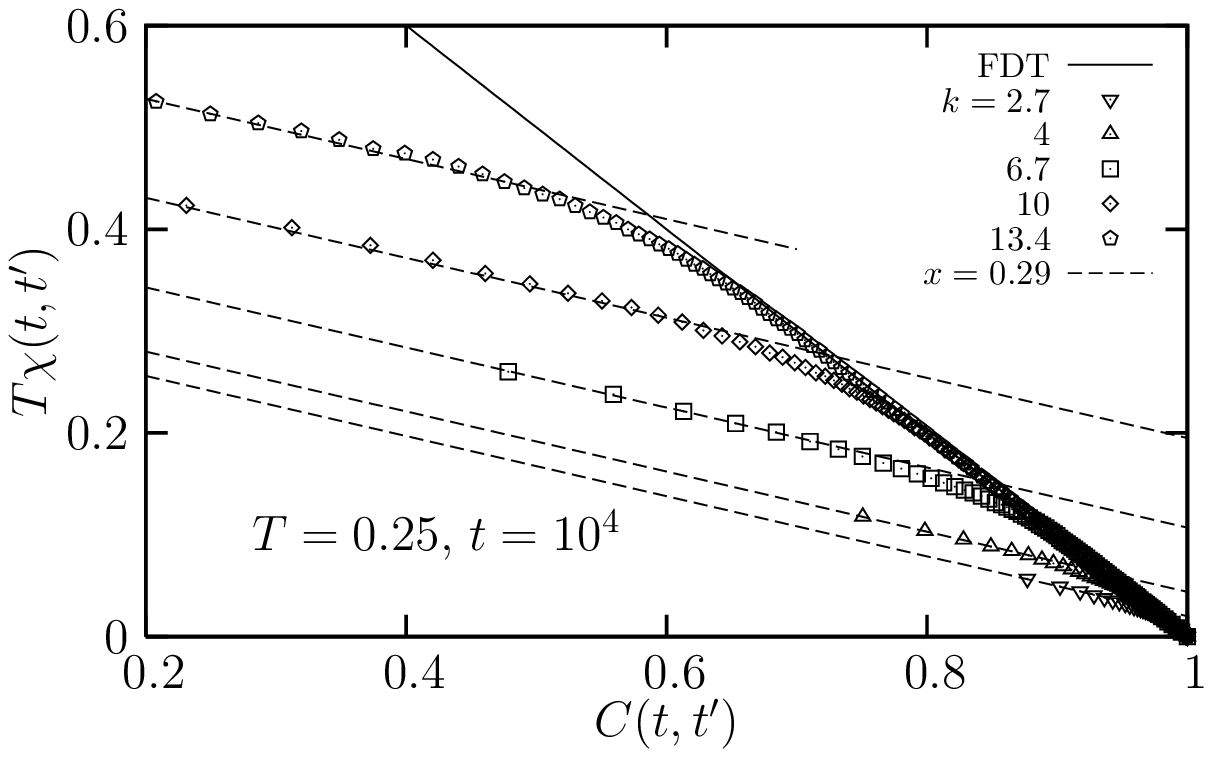,width=7.5cm}
\psfig{file=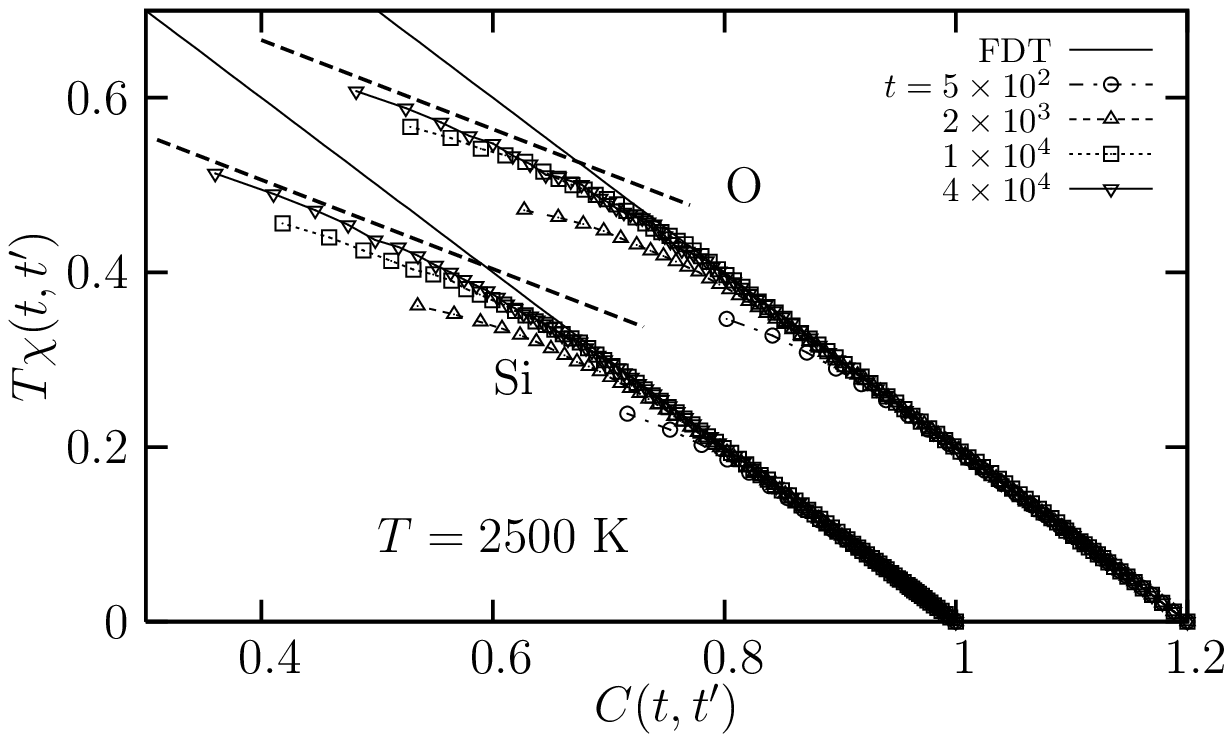,width=7.5cm}
\psfig{file=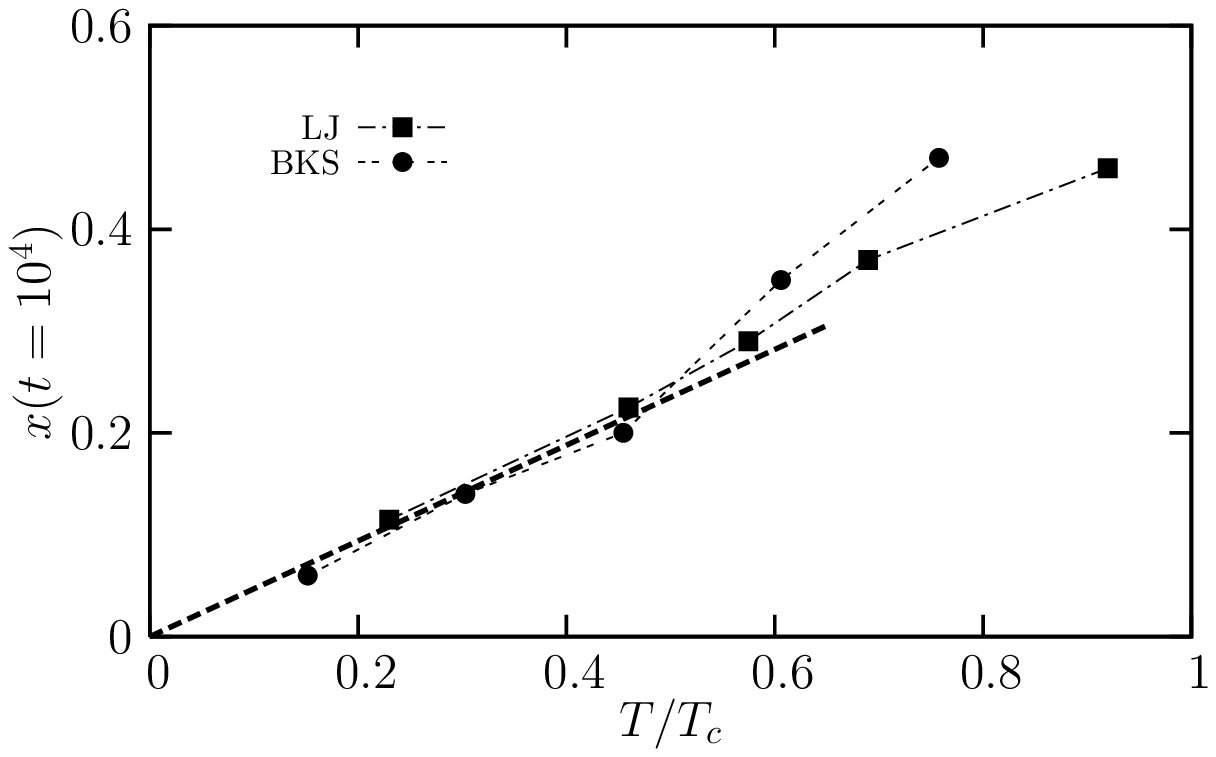,width=7.5cm}
\caption{\label{fdtT025}
Top: FD plots for fixed $T$ and $t$ in the LJ system 
and different wavevectors displaying the same nonequilibrium value
of the FDR.
Middle: FD plots for Si and O (horizontally shifted by 0.2)
in BKS for fixed $T$, $k=2.7$\AA$^{-1}$, and various $t$. For 
$t=4.10^4$, the FDR $x=0.51$ fits both sets of data.
Bottom: Temperature dependence of the FDR at a single 
large time, $x(t=10^4)$, for LJ and BKS systems. 
The temperature is normalized by the 
mode-coupling temperature $T_c$. A linear behaviour 
(dashed line) is observed at low $T$.} 
\end{figure}

We have used the flexibility offered by Eq.~(\ref{ok}) 
to characterize further the properties of FDRs
in both aging liquids in Fig.~\ref{fdtT025}. 
The top panel presents evidence
that different observables share the same FDR value, obtained
by changing the wavevector used to evaluate dynamic functions.
Similar results were obtained for silica. The middle panel
shows that Si and O atoms in silica display
similar FD plots, with equal FDR values.
Again, we find similar results for the two types
of particles in the LJ mixture.
These results suggest that it is sensible to define, 
for fixed $t$, a {\it unique} FDR value $x(t)$ characterizing 
the non-equilibrium part of FD plots.
These findings are therefore compatible
with the physical idea~\cite{jorge} that slow rearrangements in aging
supercooled liquids behave as if they were
thermalized at an ``effective temperature'' defined by
$T_{\rm eff}(t) \equiv T / x(t)$~\cite{teff}, 
with $T_{\rm eff}(t)>T$ in the two investigated systems.
Our data indicate that $T_{\rm eff}(t)$ decreases very slowly with $t$.
Finally, the bottom panel shows the temperature dependence 
of the FDR measured at a single large time, $x(t=10^4)$.
To compare both liquids we have to normalize the temperature
by some temperature scale. We choose the 
``mode-coupling'' temperature [$T_c=0.435$ (LJ)
and $T_c=3300$~K (BKS)] because 
equilibration is numerically difficult below $T_c$
and aging effects can be detected.
Remarkably, we find that FDRs in the two liquids display
a very similar temperature dependence, 
$x \approx 0.47 T/T_c$ at small $T$. This confirms that both
fragile (LJ) and strong (BKS silica) glass-formers studied in this work 
display similar aging properties. 

We have introduced a new technique
to efficiently measure linear susceptibilities 
in molecular simulations which only uses unperturbed
trajectories to evaluate response functions and outperforms
subtraction techniques in Monte-Carlo simulations.
Applied to aging supercooled liquids, 
the technique allowed us to report the first unbiased
numerical estimates of FDRs in aging molecular liquids, and 
to extend its determination to a wide range of times, 
temperatures, and observables. We 
showed that previous analysis 
quantitatively underestimated FDT violations in
LJ systems, while our results for silica are in qualitative 
disagreements with earlier results.

I thank J.-L. Barrat who suggested to reconsider the aging regime of 
BKS silica and followed this work, and
R.L. Jack and W. Kob for useful discussions and remarks on the manuscript.

\end{document}